\documentclass[final,5p,times,twocolumn,num,sort,compress]{elsarticle}
\usepackage{natbib}
\usepackage{amssymb}
\usepackage{lipsum}
\usepackage{feynmp}
\usepackage{amsmath}
\usepackage{epsfig}
\usepackage{graphicx}
\usepackage{subfigure}
\usepackage{bbm}
\usepackage{bm}
\usepackage{color}
\usepackage{threeparttable}

\journal{Physics Letters B}

\begin{document}

\begin{frontmatter}



\title{Rapid cooling of the Cassiopeia A neutron star due to superfluid quantum
criticality}


\author[addr1,addr2]{Hao-Fu Zhu}
\ead{darwinz@ustc.edu.cn}
%\orcid{0000-0001-9727-7652}
\affiliation[addr1]{organization={Department of Astronomy, University of Science and Technology of China},
            city={Hefei, Anhui},
            postcode={230026},
            country={China}}
\affiliation[addr2]{organization={School of Astronomy and Space Science, University of Science and Technology of China},
            city={Hefei, Anhui},
            postcode={230026},
            country={China}}

\author[addr3,addr4]{Guo-Zhu Liu\cormark[1]}
\ead{gzliu@ustc.edu.cn}
%\orcid{0000-0003-2144-1019}
\affiliation[addr3]{organization={Department of Modern Physics, University of Science and Technology of China},
            city={Hefei, Anhui},
            postcode={230026},
            country={China}}
\affiliation[addr4]{organization={Anhui Center for fundamental sciences in theoretical physics, University of Science and Technology of China},
            city={Hefei, Anhui},
            postcode={230026},
            country={China}}

\author[addr1,addr2]{Xufen Wu}
\ead{xufenwu@ustc.edu.cn}
%\orcid{0000-0002-1378-8082}

\cortext[1]{Corresponding author.}

\begin{abstract}
The rapid cooling of the neutron star in Cassiopeia A is speculated to arise from an enhanced neutrino emission caused by the onset of $^3P_2$-wave neutron superfluidity in the core. However, the neutrino emissivity due to Cooper-pair breaking and formation is in tension with the requirements for explaining the observed cooling rate. Here, we show that such a rapid cooling can be explained once the non-Fermi liquid behavior of the non-superfluid neutron liquid induced by superfluid quantum criticality is included into the theoretical description of neutron star cooling, without assuming the existence of additional energy loss processes. Our results indicate that the neutron star in Cassiopeia A remains in the thermal relaxation stage, which is greatly prolonged by the non-Fermi liquid behavior. The good agreement between our theoretical results and recent observational cooling data points to the pivotal role played by superfluid quantum criticality in neutron stars.
\end{abstract}



\begin{keyword}
%% keywords here, in the form: keyword \sep keyword, up to a maximum of 6 keywords
neutron star \sep superfluid quantum criticality \sep non-Fermi liquid \sep Cassiopeia A



\end{keyword}

\end{frontmatter}

%\tableofcontents



\section{Introduction}
Exploring the thermal evolution of neutron stars (NSs) \citep{Yakovlev01, Page06a, Potekhin15, Tsuruta23} is of paramount importance in astrophysics as it provides crucial constraints on many quantities ranging from the equation of state (EOS) of dense matter \citep{Lattimer16, Burgio21} to the critical temperatures of neutron superfluidity and proton superconductivity \citep{Sedrakian19, Page}. The Center Compact Object (CCO) in the supernova remnant Cassiopeia A (Cas A), which is first observed by the \emph{Chandra X-ray Observatory} in 1999 \citep{Hughes00}, exhibits a distinct rapid cooling rate from many other NSs and thus has been extensively investigated. This CCO is one of the youngest NS ($\approx340$ yr old) \citep{Fesen06} and its observed X-ray spectrum is well described by a model of a carbon atmosphere with weakly magnetized ($B \leq 10^{11}$ G) \citep{Ho09}. Based on the data acquired from the \emph{Chandra} Advanced CCD Imaging Spectrometer (ACIS), its surface temperature $T_{e}$ had been observed \citep{Ho10} to decrease by 4$\%$, from $2.12\times 10^{6}$ K to $2.04\times 10^{6}$ K, during the decade of 2000-2009. Over the subsequent decade, further analysis adjusted the cooling rate from 4$\%$ to 2$\%$ \citep{Posselt18, Posselt22, Wijngaarden19, Ho21, Shternin21, Shternin23}. However, such a rapid cooling rate is still much greater than the one ($\approx 0.7\%$) predicted by the standard cooling paradigm dominated by modified Urca (MU) processes \citep{Friman79, Yakovlev95}.

A variety of non-standard scenarios \citep{Page09, Page11, Shternin11, Blaschke12, Sedrakian13, Noda13, Yang11, Bonanno14, Negreiros13, Taranto16, Leinson14, Hamaguchi18, Leinson22, Potekhin26} have been proposed to account for the exceptional rapid cooling of the Cas A NS. Among such scenarios, the minimal cooling paradigm \citep{Page11, Shternin11} is more promising than others as it offers the most natural explanation of the late-time cooling. A key assumption of this paradigm is the recent onset of spin-triplet $^3P_2$-wave neutron superfluidity in the NS core \citep{Page11, Shternin11}. At temperatures slightly below the critical temperature $T_{\mathrm{cn}}(\rho)$, which depends on the density $\rho$, the pairing gap is quite small and thermal fluctuations can readily destroy Cooper pairs. Meanwhile, the unpaired neutrons are recombined by the attractive nuclear force into pairs. The constant pair breaking and formation (PBF) \citep{Page11}, also known as Cooper-pair formation \citep{Shternin11}, enhances the neutrino emission \citep{Flowers76, Voskresensky87, Yakovlev99b}. This provides a possible source for the extra energy loss needed to yield the rapid cooling. Moreover, ${}^1S_{0}$-wave proton superconductivity is supposed in the NS core \citep{Page11,Shternin11} for early and strong suppression of MU processes, such that it cools to the temperature observed at its current age.

Nevertheless, the minimal cooling paradigm encounters a well-known tension between theory and observation \citep{Leinson10,Shternin21,Shternin23}. The neutrino emissivity of the PBF processes is formally given by $q Q_{\nu}^{\mathrm{PBF}}$, where $Q_{\nu}^{\mathrm{PBF}}$ is the emissivity without many body corrections \citep{Yakovlev01} and $q$ represents a suppression factor that incorporates such corrections \citep{Leinson06,Leinson10}. Theoretically, the precise value of $q$ remains uncertain. Based on vector current conservation, Leinson and P{\'e}rez found that the vector channel of the PBF processes is completely suppressed \citep{Leinson06}, yielding $q=0.76$. Subsequently, Leinson considered the axial-vector channel and showed that collective effects reduce $q$ to $0.19$ in the nonrelativistic limit \citep{Leinson10}. To explain the rapid cooling of the Cas A NS, however, a larger $q$ is required. Recently, Shternin \textit{et al.} \citep{Shternin23} demonstrated that the observed rapid cooling can only be reproduced when $q$ satisfies $0.40 < q < 2.60$. This range overlaps the theoretical estimate only in ($0.40$-$0.79$), extending well beyond the upper bound ($> 0.79$). While in-medium effects and relativistic corrections may more or less modify the $q$ value \citep{Shternin21}, currently it is unclear whether they are capable of enhancing the $q$-factor to the required large values. Such a tension challenges the validity, or at least the completeness, of the minimal cooling paradigm \citep{Shternin23}. An intriguing question worth exploring is whether the rapid cooling observed in the Cas A NS can be adequately reconciled with small $q$-factor values through the inclusion of some previously neglected physical effect.

In this paper, we propose that the non-Fermi liquid (NFL) behavior of dense neutron matter induced by superfluid quantum criticality plays a unique role in the thermal evolution of NSs \citep{Zhu25} and should be carefully taken into account. The NFL behavior leads to $\propto T\ln T$ type corrections to the neutrons' specific heat $C_{\mathrm{n}}(T)$ and the total neutrino emissivity $Q_{\nu}(T)$ \citep{Zhu25}. We find that, once such singular corrections are included into the theoretical description of NS cooling, recent observational data of the effective surface temperature $T_{e}^{\infty}$ of the Cas A NS over the past 20 years can be reproduced with $q$ values within the theoretical range, down to the minimum value ($q$=$0.19$), without resort to direct Urca (DU) processes or other extra particle emission processes. Our analysis shows that this NS is currently still in the thermal relaxation stage, since the thermal relaxation time $t_{w}$ is drastically extended by quantum criticality. The maximal values of superconducting and superfluid critical temperatures, dubbed $T_{\mathrm{Cp}}$ and $T_{\mathrm{Cn}}$, respectively, can be constrained from our results.

\section{Superfluid quantum criticality}

Quantum criticality is one of the central concepts of condensed matter physics \citep{Hertz,Sachdev00,Vojta03,Chubukov03AFMQCP,Coleman05,Loehneysen07,Sachdevbook,Grover14,Pan18,Liu19,Jaoui22}. It comprises all the striking quantum many-body effects emerging near the quantum critical point (QCP) that separates the disordered (symmetric) and ordered (symmetry broken) phases. Generically, each continuous quantum phase transition driven by symmetry breaking has its own quantum criticality. Indeed, a variety of unusual properties of many condensed matter systems are believed to be caused by some type (e.g., antiferromagnetic, nematic, excitonic) of quantum criticality \citep{Vojta03, Chubukov03AFMQCP, Coleman05, Loehneysen07, Sachdevbook, Grover14, Pan18, Liu19, Jaoui22}. Especially, superconducting quantum criticality is found to exhibit intriguing behaviors in certain Dirac/Weyl fermion systems \citep{Grover14,Pan18,Liu19}. It is interesting to explore whether similar phenomena occur in NSs.

In the crust and core regions of an NS, the typical Fermi energy is on the order of $E^{}_{\mathrm{F}} \sim (10^{12}$-$10^{13})~\mathrm{K}$, while the characteristic temperature is much lower, with $k^{}_{\mathrm{B}}T < 10^{11}~\mathrm{K}$. Given that $E^{}_{\mathrm{F}}$ is significantly greater than $k^{}_{\mathrm{B}}T$, quantum fluctuations dominate over thermal fluctuations. This implies that quantum criticality is likely to play a vital role in NSs, as supported by the generic analysis presented in the literature \citep{Hertz, Sachdev00, Vojta03, Chubukov03AFMQCP, Coleman05, Loehneysen07, Sachdevbook}.

\begin{figure*}[htbp]
\centering
\includegraphics[width=5.3in]{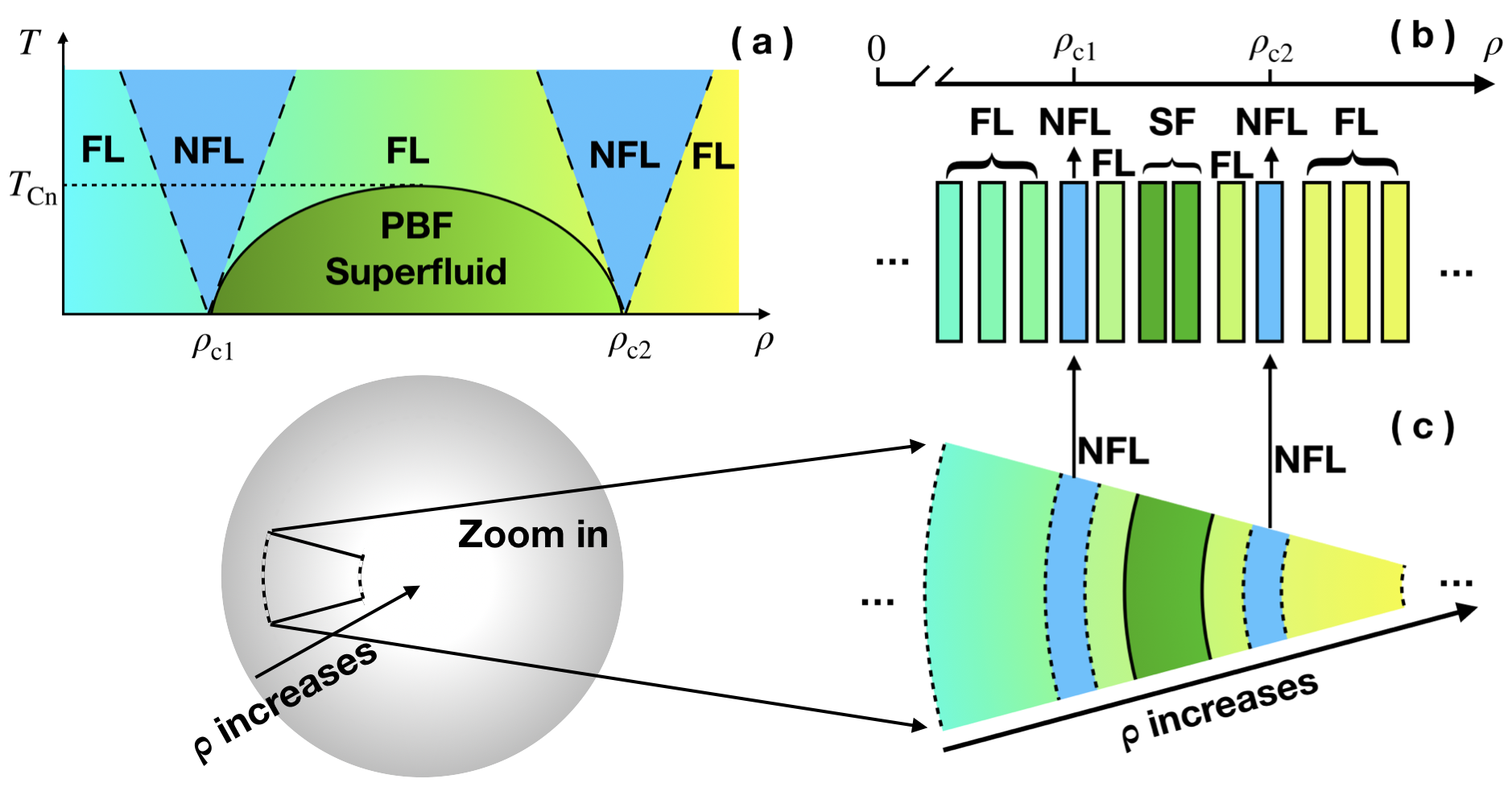}
\caption{(a): A schematic global phase diagram of neutron matter on $T$-$\rho$ plane. Here, the superfluid state specifically refers to $^1S_0$-wave neutron superfluidity. Other possible pairing states, such as $^3P_2$-wave neutron superfluidity and $^1S_0$-wave proton superconductivity, can be analyzed similarly, but for simplicity their presence is not displayed. Superfluid quantum criticality leads to NFL behavior in two broad $V$-shaped quantum critical regions. (b): An illustration of the evolution of uniform neutron matter as the nuclear density $\rho$ varies in the absence of gravity. In this panel, the temperature is finite and lower than $T_{\mathrm{Cn}}$. At this fixed temperature, increasing $\rho$ causes the uniform neutron matter to transition from a FL to a superfluid (SF), and then back to a FL. For specific values of $\rho$, the neutron matter resides in a particular state. Notably, it exhibits NFL behavior when $\rho$ is near the critical densities $\rho_{c1}$ or $\rho_{c2}$. (c): Radial cross-section of an NS depicted at the same temperature as shown in panel (b). In the radial direction pointing toward the center of the NS, the density $\rho$ is not a constant, but increases as the radius decreases. This variation in radial density leads to a stratification within the neutron matter, forming distinct layers of FL, NFL, and superfluid phases.} \label{fig:phase}
\end{figure*}

The neutron superfluid gap $\Delta_{\mathrm{n}}$ and its associated critical temperature $T_{\mathrm{cn}}$ have been extensively calculated based on the BCS theory \citep{Sedrakian19, Page}. These studies reveal a spin-singlet $^1S_0$-wave neutron superfluid in the NS crust and a spin-triplet $^3P_2$-wave neutron superfluid in the NS core. To demonstrate how quantum criticality emerges, we take the $^1S_0$-wave superfluid as an example. This superfluid state exists only within a finite density range $\rho_{c1}<\rho<\rho_{c2}$, which is schematically illustrated in Fig.~\ref{fig:phase}(a). Now consider certain amount of neutron matter. If gravity is neglected, the neutron matter is spatially uniform and its density $\rho$ is position independent. The ground state of uniform neutron matter is a Fermi liquid (FL) for $\rho < \rho_{c1}$ and $\rho > \rho_{c2}$, whereas in the range $\rho_{c1} < \rho < \rho_{c2}$ it becomes a superfluid. Near QCP $\rho_{c1}$ or $\rho_{c2}$, the uniform neutron matter exhibits quantum criticality, which is characterized by the emergence of NFL behavior \citep{Zhu25}. When the density $\rho$ is increased, zero-temperature uniform neutron matter undergoes a continuous superfluid quantum phase transition at critical densities $\rho=\rho_{c1}$ or $\rho=\rho_{c2}$. At finite temperatures, the situation becomes more complicated, as thermal fluctuations can destroy superfluidity and restore the FL state for certain densities.

For any given density $\rho$, the importance of the quantum fluctuations of superfluid order parameter $\Phi(x)$ is determined by the difference between $\rho$ and critical densities, namely $|\rho - \rho_{c1}|$ or $|\rho - \rho_{c2}|$. These fluctuations are negligible if $\rho$ is far from critical densities but become increasingly significant as $\rho$ approaches $\rho_{c1}$ or $\rho_{c2}$. In the close vicinity of $\rho_{c1}$ or $\rho_{c2}$, superfluid quantum fluctuations are sharply enhanced \citep{Zhu25}, despite that the vacuum expectation value of $\Phi(x)$ vanishes ($\langle\Phi(x) \rangle=0$). The enhanced quantum fluctuations can be described by a gapless boson mode $\phi(x)$. This boson mode is strongly coupled to the gapless neutrons excited on the Fermi surface, driving the neutron matter to exhibit NFL behavior \citep{Zhu25}. As the density departs from the critical values, superfluid quantum fluctuations do not disappear abruptly but rather decay progressively. The physical effects of these fluctuations persist over a small range around each critical density. It was demonstrated in Ref.~\citep{Zhu25} that these fluctuations can still produce NFL behavior provided the involved neutron energy $\hbar \omega$ is greater than the energy scale $E_{c}$ that is comparable to $|\rho - \rho_{c1}|$ or $|\rho - \rho_{c2}|$. At finite temperatures, thermal fluctuations also come into play and cooperate with quantum fluctuations. Their interplay broadens two zero-$T$ QCPs into two finite quantum critical regions on the $T$-$\rho$ plane, as clearly depicted in Fig.~\ref{fig:phase}(a). In these two regions, the neutron acquires an additional thermal energy $\sim k_{B}T$ due to thermal fluctuations, thus the single neutron energy is always greater than $E_{c}$. This ensures that NFL behavior exists throughout the whole quantum critical region \citep{Zhu25}. We use Fig.~\ref{fig:phase}(b) to illustrate how the uniform neutron matter evolves with growing density at a fixed finite temperature below $T_{\mathrm{Cn}}$.

In the interior of a realistic NS, the picture changes due to the presence of gravity. The equilibrium between gravity and degenerate pressure of nucleons generates a radial density gradient. In Fig.~\ref{fig:phase}(c), we show one small segment of the dense matter inside an NS along the radial direction. According to the Tolman-Oppenheimer-Volkof equation, the density $\rho(r)$ increases monotonically with decreasing radius $r$ throughout the NS. An NS is essentially constructed by stacking an infinite number of thin neutron-matter layers, each characterized by a distinct value of density, into a spherical configuration. Each layer forms a spherical shell, contributing to the overall structure of an NS. These layers correspond to the numerous states of uniform neutron matter plotted in Fig.~\ref{fig:phase}(b). Each layer has a specific radius $r$, and the two layers whose densities are close to $\rho_{c1}$ and $\rho_{c2}$ are centered near the radii $r_{c1}$ and $r_{c2}$, respectively. The FL layers, superfluid layers, and quantum critical (NFL) layers coexist within an NS. Such a coexistence is typically not realized in conventional uniform condensed-matter systems and represents a distinctive characteristic of NSs.

According to existing calculations \citep{Sedrakian19, Page}, there is a overlap between the two density regions for $^1S_0$-wave and $^3P_2$-wave superfluid phases. Within such an intermediate density region, $^1S_0$-wave and $^3P_2$-wave superfluid order parameters are both finite at low temperatures. These two orders are not independent, but compete with each other. In condensed matter physics, the competition between different long-range orders has been extensively studied in the contexts of cuprate and iron-based superconductors. As shown in Refs.~\citep{She10, Fernandes10}, two competing long-range orders (such as antiferromagnetism and superconductivity) can coexist homogeneously if their competition is weak, but are separated by a first-order transition when the competition is sufficiently strong. QCPs persist in the former case and are destroyed in the latter case. Recently, Yasui \textit{et al.} \citep{Yasui20} made a detailed analysis of the competition between $^1S_0$-wave and $^3P_2$-wave superfluid orders. In zero magnetic field, their calculations \citep{Yasui20} showed that $^1S_0$-wave superfluid excludes $^3P_2$-wave superfluid completely, which implies that the two superfluids occupy disjoint density intervals. The presence of weak magnetic fields changes the situation, allowing $^1S_0$-wave and $^3P_2$-wave superfluid orders to coexist in the overlap density region \citep{Yasui20}. In either case, $^1S_0$-wave superfluid has a well-defined upper critical density, whereas $^3P_2$-wave superfluid possesses a clear lower critical density. In another work, Sedrakian and Rau \citep{Sedrakian25} considered a distinct relation between $^1S_0$-wave and $^3P_2$-wave superfluids and found a Josephson-type current in NSs if the two superfluids are separated by a sharp interface, which might be induced by sufficiently strong ordering competition \citep{She10, Fernandes10}. Within this interface picture, no QCP survives in the intermediate-density regime. Currently, it remains unknown whether the $^1S_0$-$^3P_2$ competition is strong enough to destroy QCPs. In the present work, we will adopt the weak-competition scenario. The observation of a weak surface magnetic field ($B \leq 10^{11}$ G) in the Cas A NS \citep{Ho09} implies a comparably weak internal magnetic field, rather than the extremely strong internal field ($B \sim 10^{18}$ G) predicted by the virial theorem \citep{Lai91}. This points towards the homogeneous coexistence of $^1S_0$-wave and $^3P_2$-wave superfluid orders, in accordance with the results of Ref.~\citep{Yasui20}. In this coexisting regime, the $^1S_0$-wave gap closes at an upper critical density $\rho_{c2}$ lying within the $^3P_2$-wave superfluid domain, while the $^3P_2$-wave gap opens at a lower critical density $\rho_{c3}$ that is smaller than $\rho_{c2}$ in magnitude. Hence, the critical densities $\rho_{c2}$ and $\rho_{c3}$ correspond to two well-defined QCPs at $T=0$, which can profoundly influence the finite-temperature properties of the NS core.

The lower critical density $\rho_{c1}$ for the $^1S_0$-wave superfluid is located in the NS crust, which has a negligible effect on the NS core properties. The precise value of the upper critical density $\rho_{c4}$ for $^3P_2$-wave superfluid has not yet been determined. According to the theoretical models used in our work, $\rho_{c4}$ is roughly $4.63~\rho_{0}$ with $\rho_{0}$ denoting the nuclear saturation density, beyond the highest nuclear density reached in the core center of the Cas A NS. In the following discussions, we only consider three superfluid QCPs: $\rho_{c1}$, $\rho_{c2}$, and $\rho_{c3}$. Additionally, $^1S_0$-wave proton superconductivity is essentially independent of neutron superfluids. It has two zero-$T$ QCPs, irrespective of the existence and locations of superfluid QCPs. Superconducting quantum criticality also leads to NFL behaviors \citep{Zhu25}.

The theoretical analysis presented in \citep{Zhu25} has revealed that the NFL behavior emerging in quantum critical layers leads to qualitative changes in several physical quantities, including the neutron specific heat and neutrino emissivity, which in turn has substantial effects on the thermal evolution of NSs. Below we will show that the rapid cooling of the Cas A NS can be explained once this superfluid quantum criticality is considered.

In the absence of superfluid quantum criticality, the specific heat of neutron FL is given by
\begin{eqnarray}
C_{\mathrm{n}}^{\mathrm{FL}}(T) = \frac{k^{2}_{B}
k^{~}_{\mathrm{F}}M^{\ast}_{n}}{3\hbar^{3}}T, \label{eq:cnfl}
\end{eqnarray}
where $k^{~}_{\mathrm{F}}$ is the Fermi momentum of neutrons and $M^{\ast}_{n}$ is the effective neutron mass. Such a linear-in-$T$ expression is used in all previous studies of NS cooling. Superfluid quantum criticality generates significant NFL corrections \citep{Zhu25} to both the specific heat of neutrons and the total neutrino emissivity. The neutron specific heat \citep{Zhu25} becomes
\begin{eqnarray}
C^{\mathrm{NFL}}_{\mathrm{n}}(T) = C_{\mathrm{n}}^{\mathrm{FL}}(T) +
\frac{5k^{2}_{B}M^{\ast 2}_{n}h^{2}}{48 \hbar^{3}\pi^{2}}
T\ln\frac{T_{0}}{T}. \label{hcNFL}
\end{eqnarray}
The dimensionless parameter $h$ represents the effective strength of the coupling between gapless neutrons and superfluid quantum fluctuations. Its magnitude relies on several ingredients. As illustrated in Ref.~\citep{Zhu25}, each critical density $\rho_{ci}$ has a specific coupling constant $h_{i}$, whose value is directly related to the strength of nuclear two-body potential at $\rho_{ci}$. However, it is difficult to determine such constants since the values of critical densities are largely uncertain. In addition, the intensity of magnetic fields can affect the values of $\rho_{ci}$ and as such change the values of $h_{i}$. Nevertheless, there is currently no reliable method to estimate the influence of magnetic fields on $h_{i}$. Moreover, the width of each NFL layer depends crucially on the temperature, which also alters the physical effects of superfluid quantum critical fluctuations. The presence of superconducting QCPs makes the situation more complicated. Given the technical difficulty of calculating these coupling constants at a microscopic level, here we treat $h$ as an adjusting parameter to capture the NFL behaviors caused by all quantum critical phenomena. The combined effects of the originally complicated functions $h_{i}(\rho, T, B)$ are now captured by a single constant $h$ after averaging over the variables $\rho$, $T$, and $B$, namely
\begin{eqnarray}
\langle h_{i}(\rho, T, B)\rangle_{(\rho, T, B)} \rightarrow h. \label{averageh}
\end{eqnarray}
The parameter $h$ encapsulates not only the spatial extents of NFL-behavior layers: $\sim (C_{\mathrm{n}}^{\mathrm{NFL}}-C_{\mathrm{n}}^{\mathrm{FL}})/C_{\mathrm{n}}^{\mathrm{NFL}}$ but also the NS's internal temperature and internal magnetic field. Constraining the value of $h$ using the observed cooling data from low- to medium-mass NSs provides a way to assess whether the $^1S_0$-wave and $^3P_2$-wave superfluid orders coexist in their interiors. The symbol $T_{0}$ denotes a certain characteristic high-temperature scale of the NS, which can be defined as its internal temperature shortly after the formation in a core-collapse supernova. It is usually believed that the internal temperature of an NS can be up to $\sim 10^{11}~\mathrm{K}$ \citep{Burrows86, Fisher10, Hudepohl10}. We emphasize that the resultant cooling curves are insensitive to the precise value of $T_{0}$ within its plausible range ($10^{10}$-$10^{11}~\mathrm{K}$). Therefore, for practical purposes in fitting the observed cooling data of the Cas A NS, we use $h$ as the sole adjustable parameter while keeping $T_{0}$ fixed at the highest value.

The neutrino emissivity $Q^\mathrm{NFL}_{\nu}$ from MU processes and bremsstrahlung processes acquires a $T\ln\left(T_{0}/T\right)$ like logarithmic dependence on $T$ due to NFL corrections. The expression of $Q^\mathrm{NFL}_{\nu}$ is given in \citep{Zhu25}.

At the beginning of NS cooling history, the internal temperature $T=T_0$ and the logarithmic correction is absent since $\ln(T_0/T) = 0$. As time goes on, $T$ continues to decrease. During this process, the logarithmic correction is getting enhanced. Therefore, the NFL corrections affect the thermal evolution throughout almost the entire life of an NS. Their influence should be incorporated into the theoretical calculations of the NS cooling rate.

We use the publicly available $\texttt{NSCool}$ code package developed by Page \citep{Page16} to simulate the short-term cooling trajectory of the Cas A NS. We insert the NFL-behavior corrected quantities \citep{Zhu25}, including $C^{\mathrm{NFL}}_{\mathrm{n}}(T)$ and $Q^\mathrm{NFL}_{\nu}(T)$, into the full general-relativistic energy balance equation and the energy transport equation \citep{Thorne77}. For instance, the heat balance equation is modified to become
\begin{eqnarray}
C^\mathrm{NFL}_{\mathrm{v}}\frac{d T}{d t} = -L^\mathrm{NFL}_{\nu} -
L_{\gamma},\label{eq:heatbalancerenormalized}
\end{eqnarray}
where $L^{\mathrm{NFL}}_{\nu}=\int_{V} Q^{\mathrm{NFL}}_{\nu} dV$ is the neutrino luminosity and $L_{\gamma}$ is the surface photon luminosity. The total heat capacity of the NS core is computed as $C^{\mathrm{NFL}}_{\mathrm{v}} = \int_{V}\left(C^{\mathrm{NFL}}_{\mathrm{n}} + C_{\mathrm{p}}^{\mathrm{FL}} + C_{\mathrm{e}}^{\mathrm{FL}}\right) dV$, where $C^{\mathrm{NFL}}_{\mathrm{n}}$ is given by Eq.~(\ref{hcNFL}) and $C_{\mathrm{p}}^{\mathrm{FL}}$ and $C_{\mathrm{e}}^{\mathrm{FL}}$ are the specific heat of protons and electrons, respectively.

\begin{figure}[htbp]
\centering
\includegraphics[width=3.9in]{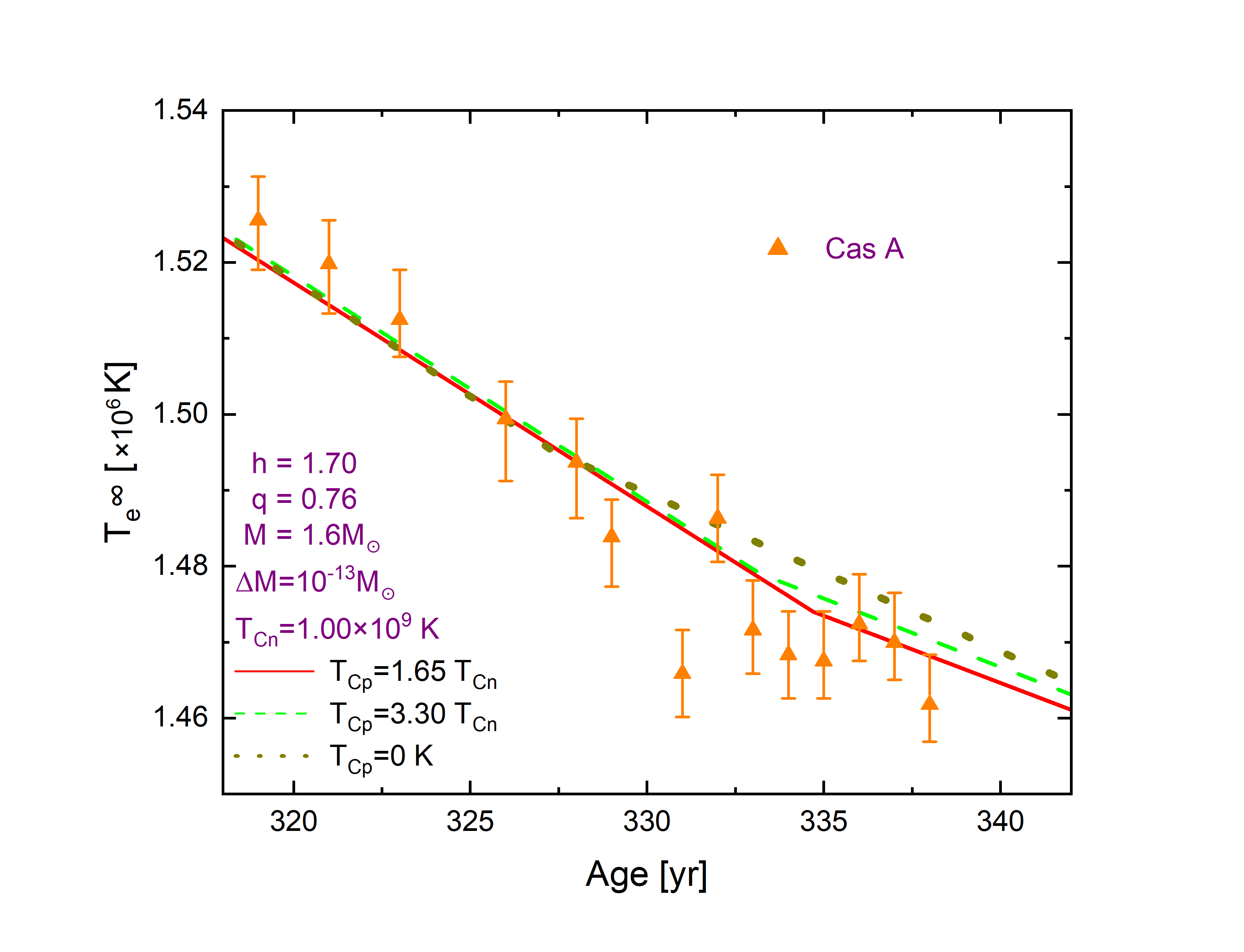}
\caption{Short-term cooling curves obtained in the presence of quantum criticality. The best fit to observations is obtained for $T_{\mathrm{Cp}}=1.65\times 10^{9}~\mathrm{K}$.}
\label{fig:shorttermcooling}
\end{figure}

As shown in the lower panel of Fig.~\ref{fig:phase}, NFL behavior emerges in two layers around $\rho_{c1}$ and $\rho_{c2}$. The layer thickness depends on $T_{e}$, or, equivalently, the age. As an NS gets colder and older, the thickness of each NFL layer is reduced, but the intensity of the logarithmic correction $\ln\left(T_{0}/T\right)$ per unit volume is amplified. The NFL and superfluid layers are separated by intermediate normal FL layers, where $C_{\mathrm{n}}^{\mathrm{FL}}(T)\sim T$. In the superfluid layer, the specific heat is strongly suppressed by the pairing gap, whereas the neutrino emission is enhanced by PBF processes \citep{Page09, Shternin11, Shternin21, Leinson22, Shternin23}. The $q$-factor enters into the corresponding neutrino emissivity $Q_{v}^{\mathrm{PBF}}$.

{It should be emphasized that $q$ and $h$ have distinct physical origins and operate in different regions of the NS interior. The factor $q$ measures the neutrino emissivity of the PBF processes and thus is nonzero only in superfluid layers. In comparison, $h$ embodies the impact of NFL behavior emerging solely in non-superfluid layers. Since the NFL behavior contribute little to the many body corrections encoded in $q$, there is no microscopic relationship between $h$ and $q$. In practice, however, the values of $q$ and $h$ are both determined by fitting the cooling observations of the Cas A NS. Consequently, the fitted values of $h$ inevitably vary as $q$ takes on different values.}

\section{Results}

The short-term cooling curves determined from our calculations are plotted in Fig.~\ref{fig:shorttermcooling}, in comparison with the cooling data of the Cas A NS (with constant X-ray absorption) extracted from the \emph{Chandra} ACIS-S Graded spectra \citep{Ho21}. Our calculations are conducted based on the Akmal-Pandharipande-Ravenhall (APR) EOS \citep{Akaml98} in the nuclear core with a carbon envelope $\Delta M=10^{-13}M_{\odot}$ \citep{Potekhin97}. The NS mass is $M = 1.6M_{\odot}$, with a central density of approximately $3.75~\rho_{0}$, which lies below the threshold for DU processes to be triggered by APR EOS, and the radius is $R = 11.35~$km. The observational cooling data of the Cas A NS (with $1\sigma$ errors) are taken from Ref.~\citep{Ho21}. Assuming that $T^2_e R$ is a constant at a fixed distance from the NS and accounting for the redshift effect, the temperature $T^\infty_e$ of observational data has been recalculated. We adopt the neutron $^1S_0$-wave superfluid model as the ``SFB'' model \citep{Page04}, which is characterized by a maximum critical temperature of approximately $5.0 \times 10^{9}~\mathrm{K}$. The lower critical density of this superfluid model is $\rho_{c1} \approx 0.00021 ~\rho_{0}$, while the upper critical density is $\rho_{c2} \approx 0.58 ~\rho_{0}$. If the factor $q$ is fixed at the upper bound of the theoretical range ($q=0.76$) \citep{Page09, Page11}, a good fit of the cooling data is achieved by choosing $h=1.70$ and $T_{\mathrm{Cn}}=1.00\times 10^{9}$~K as the ``a'' model \citep{Page04}, with the lower critical density $\rho_{c3} \approx 0.21 ~\rho_{0}$ and the upper critical density $\rho_{c4} \approx 4.63 ~\rho_{0}$. In the superfluid models considered here, the critical temperatures of them intersect at approximately $5.00\times 10^{8}$~K. For an NS of mass $M = 1.6\,M_{\odot}$, the central density is insufficient to reach the upper critical density $\rho_{c4}$ for $^3P_2$-wave superfluid phase. According to the APR EOS, such a upper critical density is attained only for NS masses above $\sim 1.9\,M_{\odot}$ \citep{Page04}, a regime disfavored by present mass constraints on the Cas A NS. We consider three different values of $T_{\mathrm{Cp}}$ as the ``T'' model \citep{Page04}, including $T_{\mathrm{Cp}} = 0~\mathrm{K},1.65\times 10^{9}~\mathrm{K},3.30\times 10^{9}~\mathrm{K}$. Setting $T_{\mathrm{Cp}}=1.65 \times 10^{9}$~K gives rise to an optimal agreement between theory and observations. Note that this $T_{\mathrm{Cp}}$ is slightly higher than $T_{\mathrm{Cn}}$, implying that the proton superconductivity is formed a little earlier than the neutron superfluid transition.

The value of $h$ may be used to constrain the nature of superfluid states and the magnetic field strength inside the Cas A NS. The NFL behavior emerging near $\rho_{c1}$ mainly affects the crust. However, the cooling rate is governed primarily by the heat capacity and neutrino emissivity of the core. To explain the observed rapid cooling of the Cas A NS, the presence of NFL effects associated with $\rho_{c2}$ and $\rho_{c3}$ in the core is required. The good agreement between our cooling curves and the cooling data of the Cas A NS favors the scenario \citep{Yasui20} in which $^1S_0$-wave and $^3P_2$-wave superfluids coexist in the NS core region rather than being separated by a first-order interface. Combined with the results of Ref.~\citep{Yasui20}, the internal magnetic field of the Cas A NS is estimated to be $B_{\mathrm{int}} \sim (0.93\text{--}6.23) \times 10^{14}~\mathrm{G}$, which is consistent with the hidden magnetic field model \citep{Vigano12} that explains its weak observed surface field ($B \leq 10^{11}~\mathrm{G}$). These results appear to establish an empirical bridge between nuclear pairing physics and macroscopic NS observables.

\begin{figure}[htbp]
\centering
\includegraphics[width=3.9in]{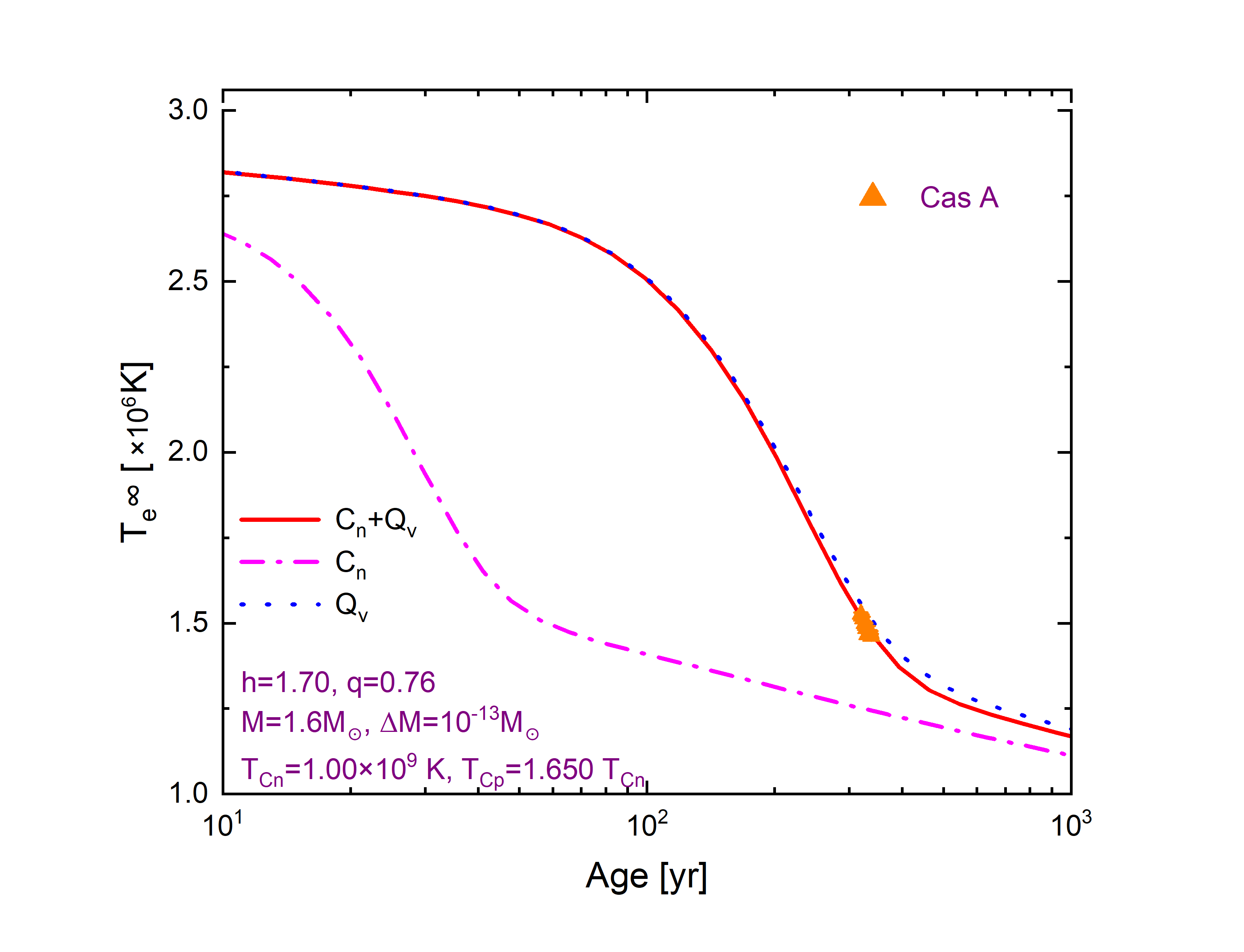}
\caption{{Long-term cooling curves of the Cas A NS with superfluid quantum criticality. The three curves show NFL corrections to: both the neutron specific heat $C_{\mathrm{n}}$ and the total neutrino emissivity $Q_{\nu}$ (solid), only $C_{\mathrm{n}}$ (dashed), and only $Q_{\nu}$ (dotted). The Cas A NS is in the thermal relaxation stage, instead of neutrino cooling stage.}} \label{fig:coolingcompare}
\end{figure}

Here we emphasize that the role of proton superconductivity differs between the minimal cooling paradigm \citep{Page11, Shternin11} and our cooling scenario. The former requires proton superconductivity to emerge at a very early stage (to suppress MU processes) and as such assumes a fairly high $T_{\mathrm{Cp}}$. The slope of the resultant cooling curve is very sensitive to the value of $T_{\mathrm{Cp}}$ \citep{Page11}. A good fit to the short-term cooling data of the Cas A NS is achieved only if $T_{\mathrm{Cp}}$ takes a special value \citep{Page11}. Such a fine tuning is not needed in our calculations, because the slope of our cooling curve is dominated by superfluid quantum criticality and is weakly modified as $T_{\mathrm{Cp}}$ varies in the range $(1.65-3.30)\times 10^{9}~\mathrm{K}$. The slope of red solid curve (i.e., optimal fitting) in Fig.~\ref{fig:shorttermcooling} is abruptly changed at the age of $334$ yr, which signals the onset of proton superconductivity. If superconductivity is absent ($T_{\mathrm{Cp}}=0~\mathrm{K}$), the cooling data can hardly be reproduced.

\begin{figure}[htbp]
\centering
\includegraphics[width=3.9in]{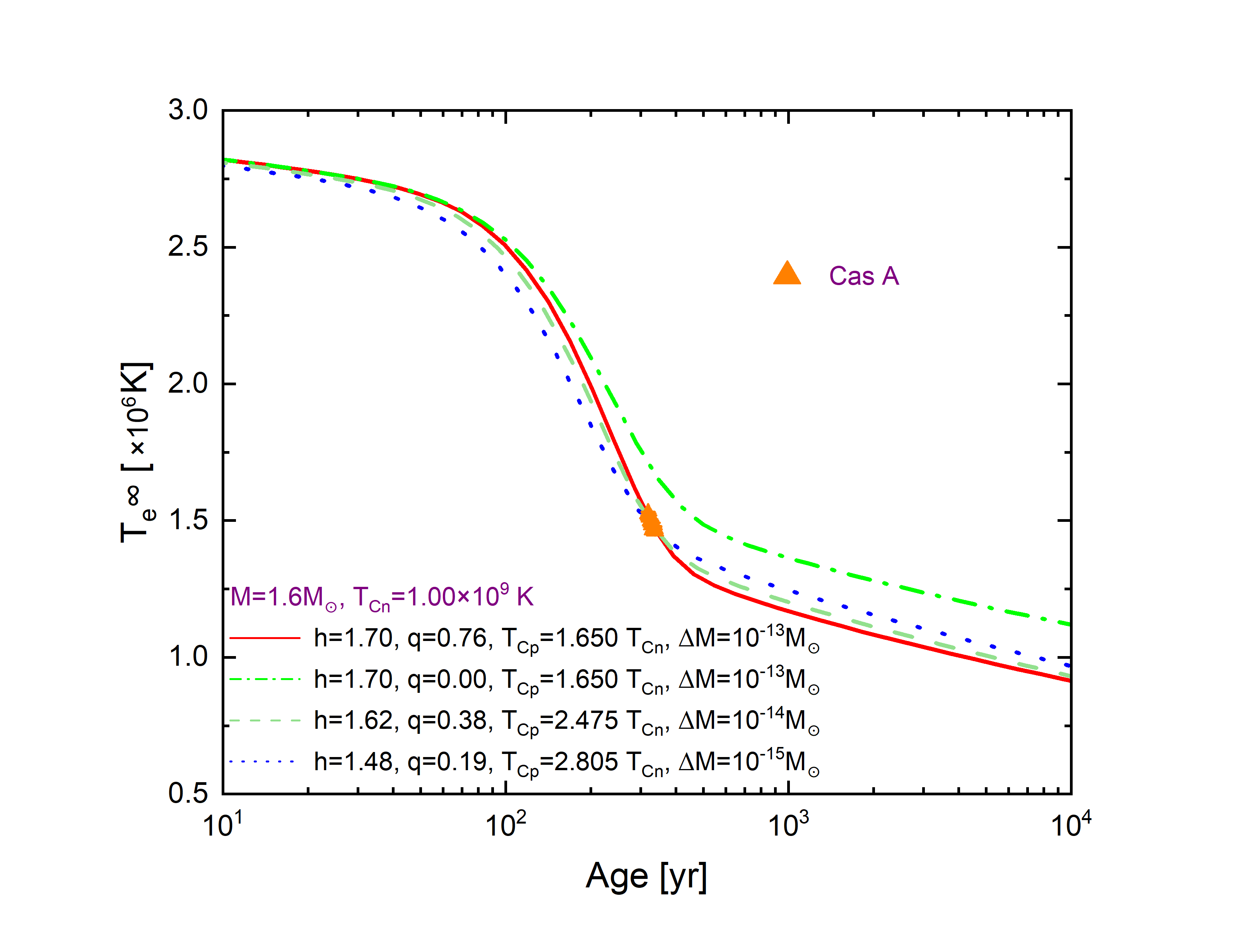}
\caption{Long-term cooling curves obtained with four sets of tuning parameters $\left(h,q,T_{\mathrm{Cp}}, \Delta M\right)$.} \label{fig:longtermcooling}
\end{figure}

The minimal cooling paradigm inferred that the Cas A NS is now in the neutrino cooling stage \citep{Page11, Shternin11}. This conclusion is drawn based on an earlier calculation \citep{Lattimer94} that estimated a thermal relaxation time $t_{w} \approx (10 - 100)$ yr. According to the long-term cooling curves presented in Fig.~\ref{fig:coolingcompare}, our cooling scenario indicates that the Cas A NS remains in the thermal relaxation stage, implying that its interior has not yet achieved an isothermal state. The difference can be understood by analyzing the following relation
\citep{Lattimer94}
\begin{eqnarray}
t_{w} \sim \frac{C^\mathrm{NFL}_{\mathrm{v}}(\Delta R)^2}{\kappa},
\label{eq:relaxationtime}
\end{eqnarray}
where $\Delta R$ is the thickness of the NS crust and $\kappa$ is the thermal conductivity. This relation shows that $t_{w}$ is proportional to the total heat capacity $C^\mathrm{NFL}_{\mathrm{v}}$. Recall that the neutrons' contribution to $C^\mathrm{NFL}_{\mathrm{v}}$ acquires a logarithmic enhancement from NFL behavior since an early time of the thermal evolution. Consequently, the value of $t_{w}$ is substantially prolonged to roughly $400$ yr. {Comparing the three curves in Fig.~\ref{fig:coolingcompare}, the NFL enhancement of the neutron specific heat is identified as the dominant effect that dramatically prolongs $t_w$, while the enhancement of the total neutrino emissivity plays a supporting role in fine-tuning the cooling slope.} The rapid cooling caused by this internal thermal non-equilibrium cooperates with the PBF processes to produce the steep slope of the cooling data shown in Fig.~\ref{fig:shorttermcooling}. Heinke and Ho \citep{Ho10} attributed the rapid cooling of the Cas A NS to its being in the thermal relaxation stage. Our cooling scenario provides a microscopic mechanism for this hypothesis. It appears that the impact of quantum criticality on the internal thermal equilibrium process is nontrivial and deserves further investigations.

\section{Discussions and conclusions}

\begin{figure}[htbp]
\centering
\includegraphics[width=3.9in]{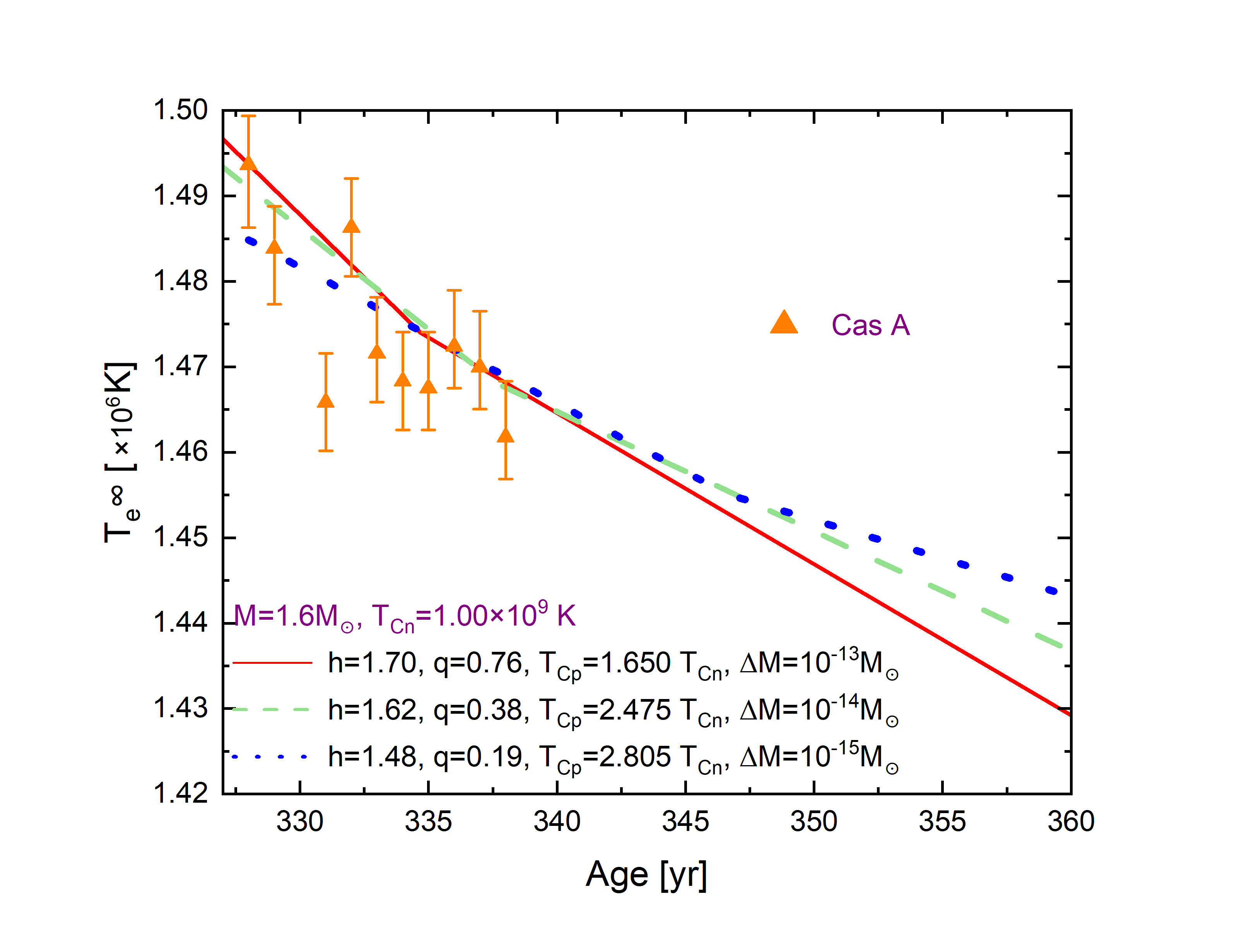}
\caption{The cooling curves of the Cas A NS from the age of $225$ yr to the age of $360$ yr. The results for the two decades after the age of $340$ yr are the prediction of our results.}
\label{fig:20yearscooling}
\end{figure}

The tension between the large $q$ ($q \geq 0.40$) demanded by minimal
cooling paradigm \citep{Shternin23} and the lower bound of theoretical range ($q=0.19$) determined by  microscopic calculations \citep{Leinson10} has sparked a fierce debate on the efficiency of minimal cooling paradigm \citep{Shternin21, Leinson22, Shternin23, Potekhin26}. To examine whether our cooling scenario still works well for small $q$, we also computed the cooling curve by choosing two small values: $q=0.38$ and $q=0.19$. Our results are compared solely with the cooling data accumulated during the decade of 2009-2019, which, due to the
remarkable advancements in both the instrument calibration and the
data analysis techniques, are more reliable than those measured from
2000 to 2009. As shown in Figs.~{\ref{fig:longtermcooling}} and
{\ref{fig:20yearscooling}}, most of the cooling data can still be
re-produced for $q=0.38$ and $q=0.19$ provided that the parameters
$h$, $T_{\mathrm{Cp}}$ and $\Delta M$ take appropriate values. {As $q$ becomes smaller, the PBF neutrino emissivity is lowered. To fit cooling data, $h$ needs to be properly reduced. Therefore, despite their distinct microscopic origins, the parameters $q$ and $h$ are effectively coupled in fitting the Cas A cooling data: a decrease in $q$ should be compensated by a decrease in $h$.}

{Within our cooling scenario, including superfluid quantum criticality into the minimal cooling paradigm provides a viable and potentially complete theory for the rapid cooling of the Cas A NS. We do not need to invoke DU processes \citep{Leinson22, Potekhin26}, which only occur under rather stringent conditions \citep{Lattimer91}, nor do we need to consider any extra exotic energy loss mechanisms (e.g., axion emission). It is worth noting that superfluid quantum criticality naturally arises whenever a zero-$T$ superfluid transition occurs at some critical density. In this sense, superfluid quantum criticality is an intrinsic feature of neutron superfluidity \citep{Zhu25}, rather than an additional assumption. With the theoretically estimated range of $q\approx (0.19$-$0.76$), restricting $h$ to the narrow interval $h\approx (1.48$-$1.70$) reproduces the observed cooling data of the Cas A NS. This offers a possible resolution to the current tension without requiring the PBF neutrino emissivity to exceed its theoretical upper bound. Notably, the inferred $h$ is specific to the Cas A NS. Its value may vary with NS mass and internal magnetic field in other NSs. Applying the same cooling scenario to other NSs will help further constrain $h$.}

Fig.~{\ref{fig:20yearscooling}} shows that the agreement between theoretical cooling curves and observed cooling data becomes less satisfactory as $q$ takes very small values, especially for the first three observational data. The agreement may be improved if the APR EOS used in our calculations is replaced by some other more realistic EOS, which is subjected to future works.

One can see from Fig.~{\ref{fig:longtermcooling}} that removing PBF processes ($q=0$) results in a much lower cooling rate during the thermal relaxation stage, which drives the cooling curve to deviate dramatically from the cooling data. This is a clear indication that PBF processes are essential in the interpretation of rapid cooling.

In addition to the ACIS data, the \emph{Chandra} High Resolution Camera (HRC) provides an alternative dataset for the cooling rate of the Cas A NS \citep{Elshamouty13, Zhao25}. Recently, Zhao \textit{et al.} reported an independent measurement of the cooling rate of the Cas A NS based on HRC data, revealing a rate of $(0.5$-$1.0)\%$ per decade \citep{Zhao25}. Zhao \textit{et al.} \citep{Zhao25} demonstrated that the HRC results of the cooling rate of the Cas A NS can be explained by the PBF-based scenario alone, even when $q=0.19$. Conversely, if the Cas A NS is indeed undergoing the rapid cooling suggested by the ACIS results, the PBF scenario alone is insufficient, and superfluid quantum criticality provides a promising complement to the PBF scenario. Currently, it is unclear which dataset, ACIS or HRC, is more reliable. The discrepancy between these two datasets highlights the need for a more in-depth analysis of potential systematic calibration issues in both the instruments \citep{Zhao25}.

The applicability of the generalized cooling scenario proposed in this work could be testified by keeping monitoring the cooling trajectory of the Cas A NS. As depicted by Fig.~\ref{fig:20yearscooling}, the cooling curves obtained at different sets of model parameters predict distinct cooling rates. Based on the cooling data to be observed in the next two decades, it is possible to verify or falsify the prediction of our cooling scenario and estimate the values of $q$, $h$, $T_{\mathrm{Cp}}$, and other parameters with a higher accuracy.

{It is worth clarifying that the thermal relaxation behavior discussed in the present work for the isolated NS in Cas A differs fundamentally from that of transiently accreting NSs \citep{Brown09}. In an isolated NS, the envelope follows the standard $T_e$–$T_b$ relation \citep{Potekhin97}, where $T_e$ is the surface temperature and $T_b$ is the temperature at the bottom of the envelope. In contrast, NS transients during outburst require a fixed $T_b$ to reproduce the observed inverted temperature profile, and relax to the $T_e$–$T_b$ relation in quiescence \citep{Brown09}. Moreover, the prolonged thermal relaxation time ($\sim400$ yr) in our scenario arises from the NFL corrections to the neutron specific heat in the core and inner crust, which are absent in the lower-density outer crusts of transients. These outer crusts govern the thermal relaxation of transients, leading to a characteristic timescale of hundreds of days \citep{Brown09}. Our scenario is therefore not inconsistent with existing studies of transients, as the two classes of sources operate under distinct envelope models and physical timescales. A detailed exploration of possible NFL effects in NS transients is beyond the present scope but merits future investigation.}

\section*{Data availability}

Data will be made available on request.

\section*{Declaration of competing interest}

The authors declare that they have no known competing financial interests or personal relationships that could have appeared to influence the work reported in this paper.

\section*{Acknowledgements}

The authors acknowledge the financial support by the National Natural Science Foundation of China under grants 12433002 and 12573018. GZL is also supported by Anhui Natural Science Foundation under grants 2508085MA003 and 2208085MA11. HFZ is also supported by China Postdoctoral Science Foundation under Grant Number 2025M783436. HFZ and XW express their gratitude to Prof. Yipeng Jing's Academician Workstation for the invaluable support and contributions and also thank the Cyrus Chun Ying Tang Foundations and the 111 Project for Observational and Theoretical Research on Dark Matter and Dark Energy (B23042). The numerical calculations in this paper have been done on the supercomputing system in the Supercomputing Center of University of Science and Technology of China.
\bibliographystyle{elsarticle-num} 
\bibliography{Casaplb}

@ARTICLE{Lattimer91,
       author = {{Lattimer}, James M. and {Pethick}, C.~J. and {Prakash}, Madappa and {Haensel}, Pawel},
        title = "{Direct URCA process in neutron stars}",
      journal = {Phys. Rev. Lett.},
         year = 1991,
        month = may,
       volume = {66},
       number = {21},
        pages = {2701-2704},
      doi = {10.1103/PhysRevLett.66.2701},
       adsurl = {https://ui.adsabs.harvard.edu/abs/1991PhRvL..66.2701L}
}

@ARTICLE{Page11,
       author = {{Page}, Dany and {Prakash}, Madappa and {Lattimer}, James M. and {Steiner}, Andrew W.},
        title = "{Rapid Cooling of the Neutron Star in Cassiopeia A Triggered by Neutron Superfluidity in Dense Matter}",
      journal = {Phys. Rev. Lett.},
         year = 2011,
        month = feb,
       volume = {106},
       number = {8},
          eid = {081101},
        pages = {081101},
         doi = {10.1103/PhysRevLett.106.081101},
archivePrefix = {arXiv},
       eprint = {1011.6142},
 primaryClass = {astro-ph.HE},
       adsurl = {https://ui.adsabs.harvard.edu/abs/2011PhRvL.106h1101P}
}

@ARTICLE{Shternin11,
       author = {{Shternin}, Peter S. and {Yakovlev}, Dmitry G. and {Heinke}, Craig O. and {Ho}, Wynn C.~G. and {Patnaude}, Daniel J.},
        title = "{Cooling neutron star in the Cassiopeia A supernova remnant: evidence for superfluidity in the core}",
      journal = {Mon. Not. R. Astron. Soc.},
         year = 2011,
        month = mar,
       volume = {412},
       number = {1},
        pages = {L108-L112},
          doi = {10.1111/j.1745-3933.2011.01015.x},
archivePrefix = {arXiv},
       eprint = {1012.0045},
 primaryClass = {astro-ph.SR},
       adsurl = {https://ui.adsabs.harvard.edu/abs/2011MNRAS.412L.108S}
}

@ARTICLE{Ho09,
       author = {{Ho}, Wynn C.~G. and {Heinke}, Craig O.},
        title = "{A neutron star with a carbon atmosphere in the Cassiopeia A supernova remnant}",
      journal = {Nature},
         year = 2009,
        month = nov,
       volume = {462},
       number = {7269},
        pages = {71-73},
         doi = {10.1038/nature08525},
archivePrefix = {arXiv},
       eprint = {0911.0672},
 primaryClass = {astro-ph.HE},
       adsurl = {https://ui.adsabs.harvard.edu/abs/2009Natur.462...71H}
}

@ARTICLE{Ho10,
       author = {{Heinke}, Craig O. and {Ho}, Wynn C.~G.},
        title = "{Direct Observation of the Cooling of the Cassiopeia A Neutron Star}",
      journal = {Astrophys. J. Lett.},
         year = 2010,
        month = aug,
       volume = {719},
       number = {2},
        pages = {L167-L171},
         doi = {10.1088/2041-8205/719/2/L167},
archivePrefix = {arXiv},
       eprint = {1007.4719},
 primaryClass = {astro-ph.HE},
       adsurl = {https://ui.adsabs.harvard.edu/abs/2010ApJ...719L.167H}
}

@ARTICLE{Posselt18,
       author = {{Posselt}, B. and {Pavlov}, G.~G.},
        title = "{Upper Limits on the Rapid Cooling of the Central Compact Object in Cas A}",
      journal = {Astrophys. J.},
         year = 2018,
        month = sep,
       volume = {864},
       number = {2},
          eid = {135},
        pages = {135},
          doi = {10.3847/1538-4357/aad7fc},
archivePrefix = {arXiv},
       eprint = {1808.00531},
 primaryClass = {astro-ph.HE},
       adsurl = {https://ui.adsabs.harvard.edu/abs/2018ApJ...864..135P}
}

@ARTICLE{Posselt22,
       author = {{Posselt}, B. and {Pavlov}, G.~G.},
        title = "{The Cooling of the Central Compact Object in Cas A from 2006 to 2020}",
      journal = {Astrophys. J.},
         year = 2022,
        month = jun,
       volume = {932},
       number = {2},
          eid = {83},
        pages = {83},
          doi = {10.3847/1538-4357/ac6dca},
archivePrefix = {arXiv},
       eprint = {2205.06552},
 primaryClass = {astro-ph.HE},
       adsurl = {https://ui.adsabs.harvard.edu/abs/2022ApJ...932...83P}
}

@ARTICLE{Yakovlev01,
       author = {{Yakovlev}, D.~G. and {Kaminker}, A.~D. and {Gnedin}, O.~Y. and {Haensel}, P.},
        title = "{Neutrino emission from neutron stars}",
      journal = {Phys. Rep.},
         year = 2001,
        month = nov,
       volume = {354},
       number = {1-2},
        pages = {1-155},
          doi = {10.1016/S0370-1573(00)00131-9},
archivePrefix = {arXiv},
       eprint = {astro-ph/0012122},
 primaryClass = {astro-ph},
       adsurl = {https://ui.adsabs.harvard.edu/abs/2001PhR...354....1Y}
}

@INCOLLECTION{Tsuruta23,
  author    = {S. Tsuruta and K. Nomoto},
  title     = {Thermal evolution of neutron stars},
  booktitle = {Handbook of Nuclear Physics},
  editor    = {I. Tanihata and H. Toki and T. Kajino},
  publisher = {Springer},
  address   = {Singapore},
  year      = {2023}
}

@article{Page06a,
   author = "Page, Dany and Reddy, Sanjay",
   title = "Dense Matter in Compact Stars: Theoretical Developments and Observational Constraints", 
   journal= "Annual Review of Nuclear and Particle Science",
   year = "2006",
   volume = "56",
   number = "Volume 56, 2006",
   pages = "327-374",
   doi = "https://doi.org/10.1146/annurev.nucl.56.080805.140600",
   publisher = "Annual Reviews",
   issn = "1545-4134",
   type = "Journal Article",}

@ARTICLE{Elshamouty13,
       author = {{Elshamouty}, K.~G. and {Heinke}, C.~O. and {Sivakoff}, G.~R. and {Ho}, W.~C.~G. and {Shternin}, P.~S. and {Yakovlev}, D.~G. and {Patnaude}, D.~J. and {David}, L.},
        title = "{Measuring the Cooling of the Neutron Star in Cassiopeia A with all Chandra X-Ray Observatory Detectors}",
      journal = {Astrophys. J.},
         year = 2013,
        month = nov,
       volume = {777},
       number = {1},
          eid = {22},
        pages = {22},
          doi = {10.1088/0004-637X/777/1/22},
archivePrefix = {arXiv},
       eprint = {1306.3387},
 primaryClass = {astro-ph.HE},
       adsurl = {https://ui.adsabs.harvard.edu/abs/2013ApJ...777...22E}
}

@misc{Zhao25,
      title={Verification of Cas A neutron star cooling rate using $\text{Chandra HRC-S}$ observations}, 
      author={Jiaqi Zhao and Craig O. Heinke and Peter S. Shternin and Wynn C. G. Ho and Dmitry D. Ofengeim and Daniel Patnaude},
      year={2025},
      eprint={2508.15161},
      archivePrefix={arXiv},
      primaryClass={astro-ph.HE},}

@ARTICLE{Lattimer94,
       author = {{Lattimer}, James M. and {van Riper}, Kenneth A. and {Prakash}, Madappa and {Prakash}, Manju},
        title = "{Rapid Cooling and the Structure of Neutron Stars}",
      journal = {Astrophys. J.},
         year = 1994,
        month = apr,
       volume = {425},
        pages = {802},
          doi = {10.1086/174025},
       adsurl = {https://ui.adsabs.harvard.edu/abs/1994ApJ...425..802L}
}

@ARTICLE{Page04,
       author = {{Page}, Dany and {Lattimer}, James M. and {Prakash}, Madappa and {Steiner}, Andrew W.},
        title = "{Minimal Cooling of Neutron Stars: A New Paradigm}",
      journal = {Astrophys. J. Suppl. Ser.},
         year = 2004,
        month = dec,
       volume = {155},
       number = {2},
        pages = {623-650},
          doi = {10.1086/424844},
archivePrefix = {arXiv},
       eprint = {astro-ph/0403657},
 primaryClass = {astro-ph},
       adsurl = {https://ui.adsabs.harvard.edu/abs/2004ApJS..155..623P}
}

@ARTICLE{Potekhin97,
       author = {{Potekhin}, A.~Y. and {Chabrier}, G. and {Yakovlev}, D.~G.},
        title = "{Internal temperatures and cooling of neutron stars with accreted envelopes.}",
      journal = {Astron. Astrophys.},
         year = 1997,
        month = jul,
       volume = {323},
        pages = {415-428},
          doi = {10.48550/arXiv.astro-ph/9706148},
archivePrefix = {arXiv},
       eprint = {astro-ph/9706148},
 primaryClass = {astro-ph},
       adsurl = {https://ui.adsabs.harvard.edu/abs/1997A&A...323..415P}
}

@ARTICLE{Thorne77,
       author = {{Thorne}, K.~S.},
        title = "{The relativistic equations of stellar structure and evolution.}",
      journal = {Astrophys. J.},
         year = 1977,
        month = mar,
       volume = {212},
        pages = {825-831},
          doi = {10.1086/155108},
       adsurl = {https://ui.adsabs.harvard.edu/abs/1977ApJ...212..825T}
}

@misc{Page16,
       author = {{Page}, Dany},
        title = "{NSCool: Neutron star cooling code}",
 howpublished = {Astrophysics Source Code Library, record ascl:1609.009},
         year = 2016,
        month = sep,
          eid = {ascl:1609.009},
       url = {https://ui.adsabs.harvard.edu/abs/2016ascl.soft09009P}
}

@ARTICLE{Jaoui22,
       author = {{Jaoui}, Alexandre and {Das}, Ipsita and {Di Battista}, Giorgio and {D{\'\i}ez-M{\'e}rida}, Jaime and {Lu}, Xiaobo and {Watanabe}, Kenji and {Taniguchi}, Takashi and {Ishizuka}, Hiroaki and {Levitov}, Leonid and {Efetov}, Dmitri K.},
        title = "{Quantum critical behaviour in magic-angle twisted bilayer graphene}",
      journal = {Nature Physics},
         year = 2022,
        month = jun,
       volume = {18},
       number = {6},
        pages = {633-638},
          doi = {10.1038/s41567-022-01556-5},
archivePrefix = {arXiv},
       eprint = {2108.07753},
 primaryClass = {cond-mat.str-el},
       adsurl = {https://ui.adsabs.harvard.edu/abs/2022NatPh..18..633J}
}

@ARTICLE{Liu19,
       author = {{Zhao}, Peng-Lu and {Liu}, Guo-Zhu},
        title = "{Absence of emergent supersymmetry at superconducting quantum critical points in Dirac and Weyl semimetals}",
      journal = {npj Quantum Materials},
         year = 2019,
        month = dec,
       volume = {4},
       number = {1},
          eid = {37},
        pages = {37},
          doi = {10.1038/s41535-019-0177-9},
       archivePrefix = {arXiv},
       eprint = {1706.02231},
       adsurl = {https://ui.adsabs.harvard.edu/abs/2019npjQM...4...37Z}
}

@ARTICLE{Pan18,
       author = {{Pan}, Xiao-Yin and {Wang}, Jing-Rong and {Liu}, Guo-Zhu},
        title = "{Quantum critical phenomena of the excitonic insulating transition in two dimensions}",
      journal = {Phys. Rev. B},
         year = 2018,
        month = sep,
       volume = {98},
       number = {11},
          eid = {115141},
        pages = {115141},
          doi = {10.1103/PhysRevB.98.115141},
archivePrefix = {arXiv},
       eprint = {1807.00452},
 primaryClass = {cond-mat.str-el},
       adsurl = {https://ui.adsabs.harvard.edu/abs/2018PhRvB..98k5141P}
}

@ARTICLE{Grover14,
       author = {{Grover}, Tarun and {Sheng}, D.~N. and {Vishwanath}, Ashvin},
        title = "{Emergent Space-Time Supersymmetry at the Boundary of a Topological Phase}",
      journal = {Science},
         year = 2014,
        month = apr,
       volume = {344},
       number = {6181},
        pages = {280-283},
          doi = {10.1126/science.1248253},
archivePrefix = {arXiv},
       eprint = {1301.7449},
 primaryClass = {cond-mat.str-el},
       adsurl = {https://ui.adsabs.harvard.edu/abs/2014Sci...344..280G}
}

@book{Sachdevbook, place={Cambridge}, edition={2}, title={Quantum Phase Transitions}, publisher={Cambridge University Press}, author={Sachdev, Subir}, year={2011}}

@ARTICLE{Loehneysen07,
       author = {{L{\"o}hneysen}, Hilbert V. and {Rosch}, Achim and {Vojta}, Matthias and {W{\"o}lfle}, Peter},
        title = "{Fermi-liquid instabilities at magnetic quantum phase transitions}",
      journal = {Reviews of Modern Physics},
         year = 2007,
        month = jul,
       volume = {79},
       number = {3},
        pages = {1015-1075},
          doi = {10.1103/RevModPhys.79.1015},
archivePrefix = {arXiv},
       eprint = {cond-mat/0606317},
 primaryClass = {cond-mat.str-el},
       adsurl = {https://ui.adsabs.harvard.edu/abs/2007RvMP...79.1015L}
}

@ARTICLE{Coleman05,
author = {{Coleman}, Piers and {Schofield}, Andrew J.},
title = "{Quantum criticality}",
journal = {Nature},
year = 2005,
month = jan,
volume = {433},
number = {7023},
pages = {226-229},
doi = {10.1038/nature03279},
archivePrefix = {arXiv},
eprint = {1301.7449},
primaryClass = {cond-mat.str-el},
adsurl = {https://ui.adsabs.harvard.edu/abs/2005Natur.433..226C }
}

@ARTICLE{Chubukov03AFMQCP,
       author = {{Abanov}, Ar. and {Chubukov}, Andrey V. and {Schmalian}, J.},
        title = "{Quantum-critical theory of the spin-fermion model and its application to cuprates: Normal state analysis}",
      journal = {Advances in Physics},
         year = 2003,
        month = mar,
       volume = {52},
       number = {3},
        pages = {119-218},
          doi = {10.1080/0001873021000057123},
      archivePrefix = {arXiv},
      eprint = {cond-mat/0107421},
       adsurl = {https://ui.adsabs.harvard.edu/abs/2003AdPhy..52..119A}
}

@article{Vojta03,
doi = {10.1088/0034-4885/66/12/R01},
year = {2003},
month = nov,
publisher = {},
volume = {66},
number = {12},
pages = {2069},
author = {Matthias Vojta},
title = {Quantum phase transitions},
journal = {Reports on Progress in Physics},}

@ARTICLE{Sachdev00,
       author = {{Sachdev}, Subir},
        title = "{Quantum Criticality: Competing Ground States in Low Dimensions}",
      journal = {Science},
         year = 2000,
        month = apr,
       volume = {288},
       number = {5465},
        pages = {475-480},
          doi = {10.1126/science.288.5465.475},
archivePrefix = {arXiv},
       eprint = {cond-mat/0009456},
 primaryClass = {cond-mat.str-el},
       adsurl = {https://ui.adsabs.harvard.edu/abs/2000Sci...288..475S}
}

@article{Hertz,
  title = {Quantum critical phenomena},
  author = {Hertz, John A.},
  journal = {Phys. Rev. B},
  volume = {14},
  issue = {3},
  pages = {1165--1184},
  numpages = {0},
  year = {1976},
  month = Aug,
  publisher = {American Physical Society},
  doi = {10.1103/PhysRevB.14.1165},}

@article{Zhu25,
    author = "Zhu, Hao-Fu and Liu, Guo-Zhu and Wang, Jing-Rong and Wu, Xufen",
    title = "{Superfluid quantum criticality and the thermal evolution of neutron stars}",
    eprint = "2408.03931",
    archivePrefix = "arXiv",
    primaryClass = "nucl-th",
    doi = "10.1103/zyvv-jv7z",
    journal = "Phys. Rev. D",
    volume = "112",
    number = "4",
    pages = "043014",
    year = "2025"
}

@ARTICLE{Leinson10,
       author = {{Leinson}, L.~B.},
        title = "{Neutrino emission from triplet pairing of neutrons in neutron stars}",
      journal = {Phys. Rev. C},
         year = 2010,
        month = feb,
       volume = {81},
       number = {2},
          eid = {025501},
        pages = {025501},
          doi = {10.1103/PhysRevC.81.025501},
archivePrefix = {arXiv},
       eprint = {0912.2164},
 primaryClass = {astro-ph.SR},
       adsurl = {https://ui.adsabs.harvard.edu/abs/2010PhRvC..81b5501L}
}

@ARTICLE{Yakovlev99b,
       author = {{Yakovlev}, D.~G. and {Kaminker}, A.~D. and {Levenfish}, K.~P.},
        title = "{Neutrino emission due to Cooper pairing of nucleons in cooling neutron stars}",
      journal = {Astron. Astrophys.},
         year = 1999,
        month = mar,
       volume = {343},
        pages = {650-660},
          doi = {10.48550/arXiv.astro-ph/9812366},
archivePrefix = {arXiv},
       eprint = {astro-ph/9812366},
 primaryClass = {astro-ph},
       adsurl = {https://ui.adsabs.harvard.edu/abs/1999A&A...343..650Y}
}

@ARTICLE{Voskresensky87,
       author = {{Voskresensky}, D.~N. and {Senatorov}, A.~V.},
        title = "{Description of a nuclear interaction in the Keldysh diagram technique and the problem of neutrino luminosity of neutron stars}",
      journal = {Soviet Journal of Nuclear Physics},
         year = 1987,
       volume = {45},
        pages = {411},
}

@article{Flowers76,
    author = "Flowers, E. and Ruderman, M. and Sutherland, P.",
    title = "{Neutrino pair emission from finite-temperature neutron superfluid and the cooling of young neutron stars}",
    doi = "10.1086/154308",
    journal = "Astrophys. J.",
    volume = "205",
    pages = "541",
    year = "1976"
}

@ARTICLE{Leinson22,
       author = {{Leinson}, Lev B.},
        title = "{Hybrid cooling of the Cassiopeia A neutron star}",
      journal = {Mon. Not. R. Astron. Soc.},
         year = 2022,
        month = apr,
       volume = {511},
       number = {4},
        pages = {5843-5848},
          doi = {10.1093/mnras/stac448},
archivePrefix = {arXiv},
       eprint = {2202.08971},
 primaryClass = {astro-ph.HE},
       adsurl = {https://ui.adsabs.harvard.edu/abs/2022MNRAS.511.5843L}
}

@ARTICLE{Hamaguchi18,
       author = {{Hamaguchi}, Koichi and {Nagata}, Natsumi and {Yanagi}, Keisuke and {Zheng}, Jiaming},
        title = "{Limit on the axion decay constant from the cooling neutron star in Cassiopeia A}",
      journal = {Phys. Rev. D},
         year = 2018,
        month = nov,
       volume = {98},
       number = {10},
          eid = {103015},
        pages = {103015},
          doi = {10.1103/PhysRevD.98.103015},
archivePrefix = {arXiv},
       eprint = {1806.07151},
 primaryClass = {hep-ph},
       adsurl = {https://ui.adsabs.harvard.edu/abs/2018PhRvD..98j3015H}
}

@ARTICLE{Leinson14,
       author = {{Leinson}, Lev B.},
        title = "{Axion mass limit from observations of the neutron star in Cassiopeia A}",
      journal = {J. Cosmol. Astropart. Phys.},
         year = 2014,
        month = aug,
       volume = {2014},
       number = {8},
        pages = {031-031},
          doi = {10.1088/1475-7516/2014/08/031},
archivePrefix = {arXiv},
       eprint = {1405.6873},
 primaryClass = {hep-ph},
       adsurl = {https://ui.adsabs.harvard.edu/abs/2014JCAP...08..031L}
}

@article{Taranto16,
    author = {Taranto, G. and Burgio, G. F. and Schulze, H.-J.},
    title = {Cassiopeia A and direct Urca cooling},
    journal = {Mon. Not. R. Astron. Soc.},
    volume = {456},
    number = {2},
    pages = {1451-1458},
    year = {2015},
    month = {12},
    issn = {0035-8711},
    doi = {10.1093/mnras/stv2756},
    eprint = {1511.04243},}

@ARTICLE{Negreiros13,
       author = {{Negreiros}, Rodrigo and {Schramm}, Stefan and {Weber}, Fridolin},
        title = "{Impact of rotation-driven particle repopulation on the thermal evolution of pulsars}",
      journal = {Phys. Lett. B},
         year = 2013,
        month = jan,
       volume = {718},
       number = {4-5},
        pages = {1176-1180},
          doi = {10.1016/j.physletb.2012.12.046},
archivePrefix = {arXiv},
       eprint = {1103.3870},
 primaryClass = {astro-ph.HE},
       adsurl = {https://ui.adsabs.harvard.edu/abs/2013PhLB..718.1176N}
}

@ARTICLE{Bonanno14,
       author = {{Bonanno}, A. and {Baldo}, M. and {Burgio}, G.~F. and {Urpin}, V.},
        title = "{The neutron star in Cassiopeia A: equation of state, superfluidity, and Joule heating}",
      journal = {Astron. Astrophys.},
         year = 2014,
        month = jan,
       volume = {561},
          eid = {L5},
        pages = {L5},
          doi = {10.1051/0004-6361/201322514},
archivePrefix = {arXiv},
       eprint = {1311.2153},
 primaryClass = {astro-ph.HE},
       adsurl = {https://ui.adsabs.harvard.edu/abs/2014A&A...561L...5B}
}

@ARTICLE{Yang11,
       author = {{Yang}, Shu-Hua and {Pi}, Chun-Mei and {Zheng}, Xiao-Ping},
        title = "{Rapid Cooling of the Neutron Star in Cassiopeia A and r-mode Damping in the Core}",
      journal = {Astrophys. J. Lett.},
         year = 2011,
        month = jul,
       volume = {735},
       number = {2},
          eid = {L29},
        pages = {L29},
          doi = {10.1088/2041-8205/735/2/L29},
archivePrefix = {arXiv},
       eprint = {1103.1092},
 primaryClass = {astro-ph.HE},
       adsurl = {https://ui.adsabs.harvard.edu/abs/2011ApJ...735L..29Y}
}

@ARTICLE{Noda13,
       author = {{Noda}, Tsuneo and {Hashimoto}, Masa-aki and {Yasutake}, Nobutoshi and {Maruyama}, Toshiki and {Tatsumi}, Toshitaka and {Fujimoto}, Masayuki},
        title = "{Cooling of Compact Stars with Color Superconducting Phase in Quark-hadron Mixed Phase}",
      journal = {Astrophys. J.},
         year = 2013,
        month = mar,
       volume = {765},
       number = {1},
          eid = {1},
        pages = {1},
          doi = {10.1088/0004-637X/765/1/1},
archivePrefix = {arXiv},
       eprint = {1109.1080},
 primaryClass = {astro-ph.SR},
       adsurl = {https://ui.adsabs.harvard.edu/abs/2013ApJ...765....1N}
}

@ARTICLE{Sedrakian13,
       author = {{Sedrakian}, Armen},
        title = "{Rapid cooling of Cassiopeia A as a phase transition in dense QCD}",
      journal = {Astron. Astrophys.},
         year = 2013,
        month = jul,
       volume = {555},
          eid = {L10},
        pages = {L10},
          doi = {10.1051/0004-6361/201321541},
archivePrefix = {arXiv},
       eprint = {1303.5380},
 primaryClass = {astro-ph.HE},
       adsurl = {https://ui.adsabs.harvard.edu/abs/2013A&A...555L..10S}
}

@ARTICLE{Blaschke12,
       author = {{Blaschke}, D. and {Grigorian}, H. and {Voskresensky}, D.~N. and {Weber}, F.},
        title = "{Cooling of the neutron star in Cassiopeia A}",
      journal = {Phys. Rev. C},
         year = 2012,
        month = feb,
       volume = {85},
       number = {2},
          eid = {022802},
        pages = {022802},
          doi = {10.1103/PhysRevC.85.022802},
archivePrefix = {arXiv},
       eprint = {1108.4125},
 primaryClass = {nucl-th},
       adsurl = {https://ui.adsabs.harvard.edu/abs/2012PhRvC..85b2802B}
}

@ARTICLE{Page09,
       author = {{Page}, Dany and {Lattimer}, James M. and {Prakash}, Madappa and {Steiner}, Andrew W.},
        title = "{Neutrino Emission from Cooper Pairs and Minimal Cooling of Neutron Stars}",
      journal = {Astrophys. J.},
         year = 2009,
        month = dec,
       volume = {707},
       number = {2},
        pages = {1131-1140},
          doi = {10.1088/0004-637X/707/2/1131},
archivePrefix = {arXiv},
       eprint = {0906.1621},
 primaryClass = {astro-ph.SR},
       adsurl = {https://ui.adsabs.harvard.edu/abs/2009ApJ...707.1131P}
}

@ARTICLE{Yakovlev95,
       author = {{Yakovlev}, D.~G. and {Levenfish}, K.~P.},
        title = "{Modified URCA process in neutron star cores}",
      journal = {Astron. Astrophys.},
         year = 1995,
        month = may,
       volume = {297},
        pages = {717},
       adsurl = {https://ui.adsabs.harvard.edu/abs/1995A&A...297..717Y}
}

@ARTICLE{Friman79,
       author = {{Friman}, B.~L. and {Maxwell}, O.~V.},
        title = "{Neutrino emissivities of neutron stars}",
      journal = {Astrophys. J.},
         year = 1979,
        month = sep,
       volume = {232},
        pages = {541-557},
          doi = {10.1086/157313},
       adsurl = {https://ui.adsabs.harvard.edu/abs/1979ApJ...232..541F}
}

@ARTICLE{Shternin23,
       author = {{Shternin}, Peter S. and {Ofengeim}, Dmitry D. and {Heinke}, Craig O. and {Ho}, Wynn C.~G.},
        title = "{Constraints on neutron star superfluidity from the cooling neutron star in Cassiopeia A using all Chandra ACIS-S observations}",
      journal = {Mon. Not. R. Astron. Soc.},
         year = 2023,
        month = jan,
       volume = {518},
       number = {2},
        pages = {2775-2793},
          doi = {10.1093/mnras/stac3226},
archivePrefix = {arXiv},
       eprint = {2211.02526},
 primaryClass = {astro-ph.HE},
       adsurl = {https://ui.adsabs.harvard.edu/abs/2023MNRAS.518.2775S}
}

@ARTICLE{Shternin21,
       author = {{Shternin}, Peter S. and {Ofengeim}, Dmitry D. and {Ho}, Wynn C.~G. and {Heinke}, Craig O. and {Wijngaarden}, M.~J.~P. and {Patnaude}, Daniel J.},
        title = "{Model-independent constraints on superfluidity from the cooling neutron star in Cassiopeia A}",
      journal = {Mon. Not. R. Astron. Soc.},
         year = 2021,
        month = sep,
       volume = {506},
       number = {1},
        pages = {709-726},
          doi = {10.1093/mnras/stab1695},
archivePrefix = {arXiv},
       eprint = {2106.05692},
 primaryClass = {astro-ph.HE},
       adsurl = {https://ui.adsabs.harvard.edu/abs/2021MNRAS.506..709S}
}

@ARTICLE{Ho21,
       author = {{Ho}, Wynn C.~G. and {Zhao}, Yue and {Heinke}, Craig O. and {Kaplan}, D.~L. and {Shternin}, Peter S. and {Wijngaarden}, M.~J.~P.},
        title = "{X-ray bounds on cooling, composition, and magnetic field of the Cassiopeia A neutron star and young central compact objects}",
      journal = {Mon. Not. R. Astron. Soc.},
         year = 2021,
        month = oct,
       volume = {506},
       number = {4},
        pages = {5015-5029},
          doi = {10.1093/mnras/stab2081},
archivePrefix = {arXiv},
       eprint = {2107.08060},
 primaryClass = {astro-ph.HE},
       adsurl = {https://ui.adsabs.harvard.edu/abs/2021MNRAS.506.5015H}
}

@ARTICLE{Wijngaarden19,
       author = {{Wijngaarden}, M.~J.~P. and {Ho}, Wynn C.~G. and {Chang}, Philip and {Heinke}, Craig O. and {Page}, Dany and {Beznogov}, Mikhail and {Patnaude}, Daniel J.},
        title = "{Diffusive nuclear burning in cooling simulations and application to new temperature data of the Cassiopeia A neutron star}",
      journal = {Mon. Not. R. Astron. Soc.},
         year = 2019,
        month = mar,
       volume = {484},
       number = {1},
        pages = {974-988},
          doi = {10.1093/mnras/stz042},
archivePrefix = {arXiv},
       eprint = {1901.01012},
 primaryClass = {astro-ph.HE},
       adsurl = {https://ui.adsabs.harvard.edu/abs/2019MNRAS.484..974W}
}

@ARTICLE{Akaml98,
       author = {{Akmal}, A. and {Pandharipande}, V.~R. and {Ravenhall}, D.~G.},
        title = "{Equation of state of nucleon matter and neutron star structure}",
      journal = {Phys. Rev. C},
         year = 1998,
        month = sep,
       volume = {58},
       number = {3},
        pages = {1804-1828},
          doi = {10.1103/PhysRevC.58.1804},
archivePrefix = {arXiv},
       eprint = {nucl-th/9804027},
 primaryClass = {nucl-th},
       adsurl = {https://ui.adsabs.harvard.edu/abs/1998PhRvC..58.1804A}
}

@ARTICLE{Fesen06,
       author = {{Fesen}, Robert A. and {Hammell}, Molly C. and {Morse}, Jon and {Chevalier}, Roger A. and {Borkowski}, Kazimierz J. and {Dopita}, Michael A. and {Gerardy}, Christopher L. and {Lawrence}, Stephen S. and {Raymond}, John C. and {van den Bergh}, Sidney},
        title = "{The Expansion Asymmetry and Age of the Cassiopeia A Supernova Remnant}",
      journal = {Astrophys. J.},
         year = 2006,
        month = jul,
       volume = {645},
       number = {1},
        pages = {283-292},
          doi = {10.1086/504254},
archivePrefix = {arXiv},
       eprint = {astro-ph/0603371},
 primaryClass = {astro-ph},
       adsurl = {https://ui.adsabs.harvard.edu/abs/2006ApJ...645..283F}
}

@ARTICLE{Hughes00,
       author = {{Hughes}, John P. and {Rakowski}, Cara E. and {Burrows}, David N. and {Slane}, Patrick O.},
        title = "{Nucleosynthesis and Mixing in Cassiopeia A}",
      journal = {Astrophys. J. Lett.},
         year = 2000,
        month = jan,
       volume = {528},
       number = {2},
        pages = {L109-L113},
          doi = {10.1086/312438},
archivePrefix = {arXiv},
       eprint = {astro-ph/9910474},
 primaryClass = {astro-ph},
       adsurl = {https://ui.adsabs.harvard.edu/abs/2000ApJ...528L.109H}
}

@INCOLLECTION{Page,
       author = {{Page}, D. and {Lattimer}, J.~M. and {Prakash}, M. and {Steiner}, A.~W.},
        title = "{Pairing and superfluidity of nucleons in neutron stars}",
    booktitle = {Novel Superfluids},
         year = 2013,
   publisher = {Oxford University Press},
     address = {Oxford, UK},
       editor = {K.H. Bennemann and J.B. Ketterson}
}

@ARTICLE{Sedrakian19,
       author = {{Sedrakian}, Armen and {Clark}, John W.},
        title = "{Superfluidity in nuclear systems and neutron stars}",
      journal = {European Physical Journal A},
         year = 2019,
        month = sep,
       volume = {55},
       number = {9},
          eid = {167},
        pages = {167},
          doi = {10.1140/epja/i2019-12863-6},
archivePrefix = {arXiv},
       eprint = {1802.00017},
 primaryClass = {nucl-th},
       adsurl = {https://ui.adsabs.harvard.edu/abs/2019EPJA...55..167S}
}

@ARTICLE{Burgio21,
       author = {{Burgio}, G.~F. and {Schulze}, H. -J. and {Vida{\~n}a}, I. and {Wei}, J. -B.},
        title = "{Neutron stars and the nuclear equation of state}",
      journal = {Progress in Particle and Nuclear Physics},
         year = 2021,
        month = sep,
       volume = {120},
          eid = {103879},
        pages = {103879},
          doi = {10.1016/j.ppnp.2021.103879},
archivePrefix = {arXiv},
       eprint = {2105.03747},
 primaryClass = {nucl-th},
       adsurl = {https://ui.adsabs.harvard.edu/abs/2021PrPNP.12003879B}
}

@ARTICLE{Lattimer16,
       author = {{Lattimer}, James M. and {Prakash}, Madappa},
        title = "{The equation of state of hot, dense matter and neutron stars}",
      journal = {Phys. Rep.},
         year = 2016,
        month = mar,
       volume = {621},
        pages = {127-164},
          doi = {10.1016/j.physrep.2015.12.005},
archivePrefix = {arXiv},
       eprint = {1512.07820},
 primaryClass = {astro-ph.SR},
       adsurl = {https://ui.adsabs.harvard.edu/abs/2016PhR...621..127L}
}

@ARTICLE{Potekhin15,
       author = {{Potekhin}, Alexander Y. and {Pons}, Jos{\'e} A. and {Page}, Dany},
        title = "{Neutron Stars{\textemdash}Cooling and Transport}",
      journal = {Space Sci. Rev.},
         year = 2015,
        month = oct,
       volume = {191},
       number = {1-4},
        pages = {239-291},
          doi = {10.1007/s11214-015-0180-9},
archivePrefix = {arXiv},
       eprint = {1507.06186},
 primaryClass = {astro-ph.HE},
       adsurl = {https://ui.adsabs.harvard.edu/abs/2015SSRv..191..239P}
}

@ARTICLE{Burrows86,
       author = {{Burrows}, A. and {Lattimer}, J.~M.},
        title = "{The Birth of Neutron Stars}",
      journal = {Astrophys. J.},
         year = 1986,
        month = aug,
       volume = {307},
        pages = {178},
          doi = {10.1086/164405},
       adsurl = {https://ui.adsabs.harvard.edu/abs/1986ApJ...307..178B}
}

@ARTICLE{Fisher10,
       author = {{Fischer}, T. and {Whitehouse}, S.~C. and {Mezzacappa}, A. and {Thielemann}, F.-K. and {Liebend{\"o}rfer}, M.},
        title = "{Protoneutron star evolution and the neutrino-driven wind in general relativistic neutrino radiation hydrodynamics simulations}",
      journal = {Astron. Astrophys.},
         year = 2010,
        month = jul,
       volume = {517},
          eid = {A80},
        pages = {A80},
          doi = {10.1051/0004-6361/200913106},
archivePrefix = {arXiv},
       eprint = {0908.1871},
 primaryClass = {astro-ph.HE},
       adsurl = {https://ui.adsabs.harvard.edu/abs/2010A&A...517A..80F}
}

@ARTICLE{Hudepohl10,
       author = {{H{\"u}depohl}, L. and {M{\"u}ller}, B. and {Janka}, H.-T. and {Marek}, A. and {Raffelt}, G.~G.},
        title = "{Neutrino Signal of Electron-Capture Supernovae from Core Collapse to Cooling}",
      journal = {Phys. Rev. Lett.},
         year = 2010,
        month = jun,
       volume = {104},
       number = {25},
          eid = {251101},
        pages = {251101},
          doi = {10.1103/PhysRevLett.104.251101},
archivePrefix = {arXiv},
       eprint = {0912.0260},
 primaryClass = {astro-ph.SR},
       adsurl = {https://ui.adsabs.harvard.edu/abs/2010PhRvL.104y1101H}
}

@ARTICLE{Yasui20,
       author = {{Yasui}, Shigehiro and {Inotani}, Daisuke and {Nitta}, Muneto},
        title = "{Coexistence phase of $^{1}${S}$_{0}$ and $^{3}${P}$_{2}$ superfluids in neutron stars}",
      journal = {Phys. Rev. C},
         year = 2020,
        month = may,
       volume = {101},
       number = {5},
          eid = {055806},
        pages = {055806},
          doi = {10.1103/PhysRevC.101.055806},
archivePrefix = {arXiv},
       eprint = {2002.05429},
 primaryClass = {nucl-th},
       adsurl = {https://ui.adsabs.harvard.edu/abs/2020PhRvC.101e5806Y}
}

@ARTICLE{Sedrakian25,
       author = {{Sedrakian}, Armen and {Rau}, Peter B.},
        title = "{Josephson currents in neutron stars}",
      journal = {Phys. Rev. D},
         year = 2025,
        month = jan,
       volume = {111},
       number = {2},
          eid = {023044},
        pages = {023044},
          doi = {10.1103/PhysRevD.111.023044},
archivePrefix = {arXiv},
       eprint = {2407.13686},
 primaryClass = {astro-ph.HE},
       adsurl = {https://ui.adsabs.harvard.edu/abs/2025PhRvD.111b3044S}
}

@article{She10,
  title = {Stability of quantum critical points in the presence of competing orders},
  author = {She, Jian-Huang and Zaanen, Jan and Bishop, Alan R. and Balatsky, Alexander V.},
  journal = {Phys. Rev. B},
  volume = {82},
  issue = {16},
  pages = {165128},
  numpages = {20},
  year = {2010},
  month = {Oct},
  publisher = {American Physical Society},
  doi = {10.1103/PhysRevB.82.165128},
  archivePrefix = {arXiv},
  eprint = {1009.1888},}

@article{Fernandes10,
  title = {Competing order and nature of the pairing state in the iron pnictides},
  author = {Fernandes, Rafael M. and Schmalian, J\"org},
  journal = {Phys. Rev. B},
  volume = {82},
  issue = {1},
  pages = {014521},
  numpages = {22},
  year = {2010},
  month = {Jul},
  publisher = {American Physical Society},
  doi = {10.1103/PhysRevB.82.014521},
  archivePrefix = {arXiv},
  eprint = {1005.2437},}

@ARTICLE{Vigano12,
       author = {{Vigan{\`o}}, D. and {Pons}, J.~A.},
        title = "{Central compact objects and the hidden magnetic field scenario}",
      journal = {Mon. Not. R. Astron. Soc.},
         year = 2012,
        month = oct,
       volume = {425},
       number = {4},
        pages = {2487-2492},
          doi = {10.1111/j.1365-2966.2012.21679.x},
archivePrefix = {arXiv},
       eprint = {1206.2014},
 primaryClass = {astro-ph.SR},
       adsurl = {https://ui.adsabs.harvard.edu/abs/2012MNRAS.425.2487V}
}

@ARTICLE{Lai91,
       author = {{Lai}, Dong and {Shapiro}, Stuart L.},
        title = "{Cold Equation of State in a Strong Magnetic Field: Effects of Inverse beta -Decay}",
      journal = {Astrophys. J.},
         year = 1991,
        month = dec,
       volume = {383},
        pages = {745},
          doi = {10.1086/170831},
       adsurl = {https://ui.adsabs.harvard.edu/abs/1991ApJ...383..745L}
}

@ARTICLE{Brown09,
       author = {{Brown}, Edward F. and {Cumming}, Andrew},
        title = "{Mapping Crustal Heating with the Cooling Light Curves of Quasi-Persistent Transients}",
      journal = {Astrophys. J.},
         year = 2009,
        month = jun,
       volume = {698},
       number = {2},
        pages = {1020-1032},
          doi = {10.1088/0004-637X/698/2/1020},
archivePrefix = {arXiv},
       eprint = {0901.3115},
 primaryClass = {astro-ph.SR},
       adsurl = {https://ui.adsabs.harvard.edu/abs/2009ApJ...698.1020B}
}

@ARTICLE{Leinson06,
       author = {{Leinson}, L.~B. and {P{\'e}rez}, A.},
        title = "{Vector current conservation and neutrino emission from singlet-paired baryons in neutron stars}",
      journal = {Physics Letters B},
         year = 2006,
        month = jul,
       volume = {638},
       number = {2-3},
        pages = {114-118},
          doi = {10.1016/j.physletb.2006.05.036},
archivePrefix = {arXiv},
       eprint = {astro-ph/0606651},
 primaryClass = {astro-ph},
       adsurl = {https://ui.adsabs.harvard.edu/abs/2006PhLB..638..114L}
}

@ARTICLE{Potekhin26,
       author = {{Potekhin}, A.~Y. and {Yakovlev}, D.~G.},
        title = "{Urca cooling of the neutron star in the Cassiopeia A supernova remnant}",
      journal = {Journal of High Energy Astrophysics},
         year = 2026,
        month = jan,
       volume = {49},
          eid = {100441},
        pages = {100441},
          doi = {10.1016/j.jheap.2025.100441},
       adsurl = {https://ui.adsabs.harvard.edu/abs/2026JHEAp..4900441P}
}



%% For example:

%% \bibitem[Aladro et al.(2015)]{Aladro15} Aladro, R., Martín, S., Riquelme, D., et al. 2015, \aas, 579, A101

%%\end{thebibliography}

\end{document}